\documentclass{appolb}
\usepackage{graphicx}
\usepackage{amsmath}
\usepackage{amsfonts}
\usepackage{subcaption}

\begin{document}

\title{Cylindrical Euler\,--\,Poisson Equation in a Chaplygin Gas Medium%
}
\author{Balázs E. Szigeti,$^{1,2}$, Imre F. Barna,$^{2,3}$, Gergely G. Barnaf\"oldi$^{1}$
\address{$^{1}$HUN-REN Wigner RCP, 29-33 Konkoly-Thege M. Str., 1121 Budapest, Hungary}
\\
\address{$^{2}$E\"otv\"os Lor\'and University, P\'azm\'any P\'eter stny. 1/A, 1117 Budapest, Hungary}
\\
}

\maketitle
\begin{abstract}
We studied the Euler\,--\,Poisson equation system in the case of cylindrical symmetry with the von Neumann\,--\,Sedov\,--\,Taylor type of self-similar {\it ansatz} and present scaling solutions. We have analyzed the scenario governed by Chaplygin’s equation of state, which has historically been studied as a unifying framework of dark fluid for dark matter and dark energy.
\end{abstract}
\section{Introduction}

The evolution of a blast wave generated by a powerful explosion has been a subject of great interest since its initial study during the '40s and '50s. The sudden release of a large amount of energy in a confined region creates a discontinuity surface, across which physical quantities such as density, velocity, and possibly temperature exhibit abrupt changes~\cite{Gouderley1942, Sedov1946}. This discontinuity surface, known as the shock front, has been extensively studied over the past few decades. A fundamental \emph{ansatz} to this problem was first introduced by von Neumann~\cite{Neumann1941}, Sedov~\cite{Sedov1959}, and Taylor~\cite{Taylor1950}, leading to what is now known as the von Neumann \,--\, Sedov\,--\,Taylor solution. These solutions exhibit self-similar behaviour at intermediate timescales, where the system's evolution is governed by scaling laws that bridge initial transients and final equilibrium states through dynamically emerging similarity variables~\cite{zeldov}. 
 
In our analysis, we studied the Euler\,--\,Poisson equation system, a fundamental tool in astrophysics governing the dynamics of self-gravitating fluids. Historically, it has crucial applications in star formation~\cite{Emden1907}, gravitational collapse, cosmological structure formation~\cite{Horedt2004} and heavy nuclear synthesis. The von Neumann\,--\, Sedov\,--\ Taylor blast wave models inspired us to construct a non-relativistic dark fluid model, which might have a new view on the evolution of the Universe~\cite{NMRAS}. 


\section{Euler\,--\,Poisson Equation System}

Since rotation can break the spherical symmetry, axial solutions can also have strong relevance. Here, we extend our previous analyses of spherical flows to the Euler\,--\,Poisson equation in a cylindrical system~\cite{Barna2022, Szigeti2023}. 
\begin{align}
    \partial_t \rho + (\partial_r \rho) u + (\partial_r u) \rho + \dfrac{u\rho}{r} &= 0 \label{eq:03A} \ , \\
    \partial_t u + (u \partial_r) u &= - \dfrac{1}{\rho} \partial_r P  - \partial_r \Phi \label{eq:03B} \ , \\
    \dfrac{1}{r} \dfrac{\mathrm{d}}{\mathrm{d}r}\left( r \partial_r \Phi \right)  & = 4 \pi \rho\label{eq:03C} \ .
\end{align}
We solved the equations by using the Neumann\,--\,Sedov\,--\,Taylor ansatz in geometrical units for the velocity field $u=u(r,t)$, the density $\rho=\rho(r,t)$ and the gravitational potential density field $\Phi=\Phi(r,t)$, 
\begin{align}
     u (r,t)  =  t^{-\alpha} f(\eta), \quad
    \rho(r,t) = t^{-\gamma} g (\eta), \ \mathrm{and} \quad
    \Phi(r,t) = t^{-\delta} h(\eta),
    \label{eq::04}
\end{align}
where $f(\eta)$, $g(\eta)$ and $h(\eta)$ are the shape functions of the reduced ordinary differential equation system with the reduced variable of $\eta = r/t^{\beta}$. We have examined the scenario described by Chaplygin's equation of state, $P(\rho) = -A\rho^{-n}$ with $A\in \mathbb{R}^+$ and $-1<n\leq 1$, describing both dark matter and dark energy as a unified dark fluid~\cite{chap_dark1,chap_dark2}. The Chaplygin gas has historically been explored as a unifying framework for dark matter and energy, offering a smooth transition from a pressureless dust-like regime to an accelerating cosmological phase~\cite{GUO2007326}. Its generalizations have been employed in studies of structure formation and modifications of the cosmic expansion history, particularly in alternative gravity and brane-world scenarios~\cite{PhysRevD.62.085023}. By substituting Eqs.~\eqref{eq::04} into the Euler–Poisson equation system~\eqref{eq:03A}-\eqref{eq:03C}, a coupled differential equation is obtained along with an underdetermined algebraic equation, which constrains the similarity exponents. 
 \begin{table}[!h]
    \centering
    \begin{tabular}{crrcr}
     \hline
     \hline
\multicolumn{1}{c}{Solution} & \multicolumn{4}{c}{Shape-functions} \\
Variant & $\alpha$ & $\beta$ & $\gamma$ & $\delta$ \\
     \hline
      (i) & 1/2 & 1/2 & $-1/(n+1)$ & 1 \\ 
      (ii) & 1 & 0 & -1 & 2 \\ 
      (iii) & 0 & 1 & 0 & 0 \\ 
    \hline
    \hline
    \end{tabular}
    \caption{Similarity exponents expressed in function of $n$. The analysis follows the methodology outlined in Ref.~\cite{Szigeti2023}.}
    \label{tab:similarity_exponents}
\end{table}
Table~\ref{tab:similarity_exponents} shows the obtained numerical values of the similarity exponents expressed in terms of the $n$ Chaplygin exponent. At the next derivation stage, the induced partial differential equations (PDEs) system is transformed into a system of ordinary differential equations (ODEs) that depends solely on the independent variable $\eta$. A comprehensive examination of the $n = -1$ case can be found in our previous work~\cite{Szigeti2023}.
\begin{align}
 f'(\eta) g(\eta) + f(\eta) g'(\eta) + \dfrac{f(\eta)g(\eta)}{\eta} & = \gamma g(\eta) + \beta \eta g'(\eta), \label{eq::2A}\\  
 -\alpha f(\eta) - \beta \eta f'(\eta)  -\eta^2 f'(\eta) + \eta f'(\eta) f(\eta) &  = -n A  g^{-(n+2)}(\eta) g'(\eta) - \eta h'(\eta),\label{eq::2B}\\
   h'(\eta) + h''(\eta) \eta & =  4 \pi 
   \eta \ g (\eta) . \label{eq::2C}
\end{align}
Physically relevant time-decaying solutions should have small non-negative $\alpha,\gamma, \delta$ exponents. In addition, the $\beta$ exponent characterizes the spreading of the solution functions in time. Regular diffusion or incompressible Navier\,--\,Stokes equations have exponents with a numerical value of $1/2$. The larger the exponents' absolute values, the more radical the dynamic variables' temporal change.
\begin{figure}[!h]
    \centering
    \begin{subfigure}{0.495\textwidth}
        \centering
        \includegraphics[width=\linewidth]{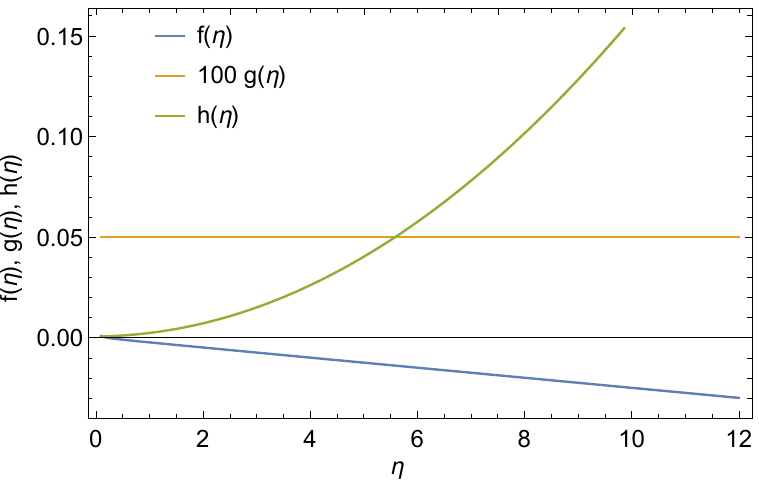}
        \label{fig:first}
    \end{subfigure}
    \hfill
    \begin{subfigure}{0.495\textwidth}
        \centering
        \includegraphics[width=\linewidth]{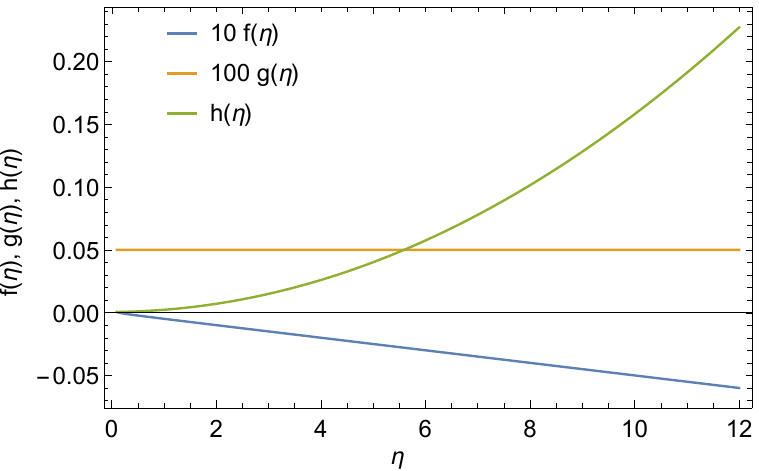}
        \label{fig:second}
    \end{subfigure}
    \begin{subfigure}{0.495\textwidth}
        \centering
        \includegraphics[width=\linewidth]{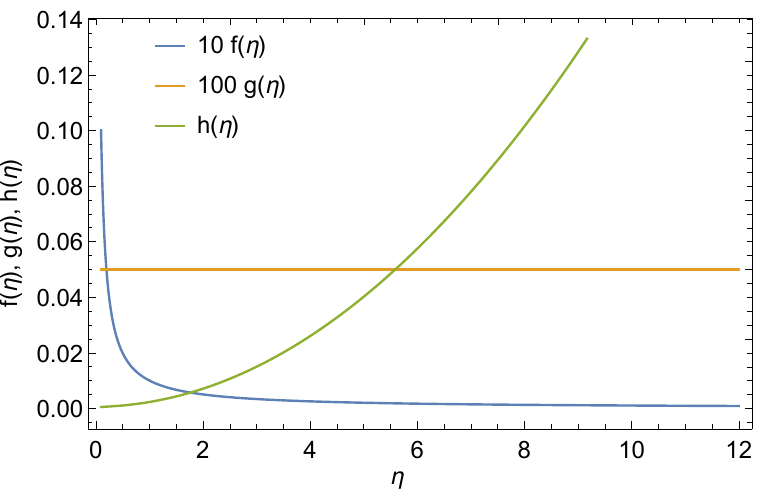}
        \label{fig:third}
    \end{subfigure}
    \caption{The shape-functions (for EoS variants (i)-(iii) from left to right and top to bottom respectively) and for $n=1$, as solutions of the reduced Euler\,--\,Poisson equation function expressed as function $\eta$ similarity variable. }
\end{figure}
\newpage
A systematic approach involves numerically solving the obtained system of ordinary differential Eqs.~\eqref{eq::2A}-\eqref{eq::2C} for a wide range of parameter sets informed by physical considerations. Figure 1 illustrates the three related shape functions $f(\eta), g(\eta)$ and $h(\eta)$ respectively, for each Chaplygin EoS variant. Our analysis revealed that only variant (iii) exhibits an expansion behaviour analogous to our previously explored linear EoS dark fluid model. The others lead to a Newtonian core-collapse scenario, as indicated by the velocity shape function  $f(\eta)$ becoming negative. In the scenario of variant (iii), one can see, that the density exhibits a constant behaviour, whereas the velocity diminishes to zero in the asymptotic limit of large time. Among the investigated scenarios, only this one aligns with the properties of the present-day Friedmannian universe.
\section{Summary} 
In this manuscript, we analysed the cylindrical Euler\,--\,Poisson hydrodynamical equation system closed with the Chapligin gas equation of state with the time-dependent self-similar {\it ansatz}. The reduced ODE system was integrated numerically the obtained fluid velocity, density and gravitation field distributions were analyzed for various Chaplygin EoS variants. Our analysis found that only variant (iii) exhibits expansion behaviour similar to the linear EoS dark fluid model, with constant density and vanishing velocity at late times, making it the sole scenario consistent with the present-day Friedmannian universe.

\section*{Acknowledgment}
We acknowledge the financial support by the Hungarian NRDIO under Contract No., 2021-4.1.2-NEMZ\_KI-2024-00058, 2024-1.2.5-TÉT-2024-00022 and Wigner Scientific Computing Laboratory (WSCLAB).

\bibliographystyle{IEEEtran}
\bibliography{dark}
\end{document}